# Application of high-spatial-resolution secondary ion mass spectrometry for nanoscale chemical mapping of lithium in an Al-Li alloy

Xu Xu [a], Chengge Jiao [b], Kexue Li [a,c], Min Hao [d], Katie. L. Moore [a,c], Timothy. L. Burnett [a,c,*], Xiaorong Zhou [a]

[a] School of Materials, University of Manchester, Manchester, M13 9PL, UK

[b] Thermo Fisher Scientific, Achtseweg Noord 5, 5651 GG, Eindhoven, the Netherlands.

[c] Photon Science Institute, University of Manchester, Manchester, M13 9PL, UK

[d] Aluminium Alloys and Magnesium Alloys Research Institute, Beijing Institute of Aeronautical Materials, Beijing, 100095, China

*Corresponding author.
Email: timothy.burnett@manchester.ac.uk

**Abstract**

High-spatial-resolution secondary ion mass spectrometry offers a method for mapping lithium at nanoscale lateral resolution. Practical implementation of this technique offers significant potential for revealing the distribution of Li in many materials with exceptional lateral resolution and elemental sensitivity. Here, two state-of-the-art methods are demonstrated on an aluminium-lithium alloy to visualise nanoscale Li-rich phases by mapping the $^7Li^+$ secondary ion. NanoSIMS 50L analysis with a radio frequency $O^-$ plasma ion source enabled visualisation of needle-shaped $T_1$ ($Al_2CuLi$) phases as small as 75 nm in width. A compact time-of-flight secondary ion mass spectrometry detector added to a focused ion beam scanning electron microscope facilitated mapping of the $T_1$ phases down to 45 nm in width using a $Ga^+$ ion beam. Correlation with high resolution electron microscopy confirms the identification of $T_1$ precipitates, their sizes and distribution observed during SIMS mapping.

**Key words**

Nanoscale mapping of lithium, high-spatial-resolution secondary ion mass spectrometry, NanoSIMS, FIB ToF-SIMS, aluminium-lithium alloy

**Main text**

The characterisation of lithium (Li) has received considerable attention from multiple disciplines in materials science in particular researchers working on the development of Li-ion battery materials and high-strength, low-density Al-Li alloys [1–9]. High-spatial-resolution chemical mapping of Li is of particular interest in the case of Li-ion battery materials for investigating Li ion transport on the electrodes with nanoscale structures [1,8,9]. It also offers excellent potential for revealing the distribution of nanoscale Li-rich precipitates in Al-Li alloys to understand their influence on mechanical and corrosion performance [4–7]. However, the analysis of Li is challenging as many techniques, such as energy dispersive X-ray spectroscopy (EDX or EDS) and X-ray photoelectron spectroscopy (XPS), have low elemental sensitivity for Li and/or insufficient spatial resolution [3,10]. While the use of electron energy loss spectroscopy (EELS) and atom probe tomography (APT) enables the characterisation of Li with high spatial resolution if carried out with great care, these techniques are only able to cover extremely small areas due to the complexity of sample preparation and small sampling volume (e.g. ~ $1 \times 10^{-2}$ µm$^2$ and ~ $5 \times 10^{-3}$ µm$^3$ for EELS mapping and APT, respectively) [11,12].

Secondary ion mass spectrometry (SIMS) is a surface analysis technique that provides detailed chemical analysis and has been used widely on solid materials [13]. Elemental, isotopic and molecular information can be obtained from the top surface of a solid sample after bombardment by incident primary ions. The sputtered secondary ions are then extracted, transported and separated by a mass spectrometer according to their mass-to-charge ratio. SIMS offers excellent elemental sensitivity, in the parts per million to parts per billion range, in combination with the ability to detect light elements, including hydrogen, as well as isotopes from areas measuring 10's -100's µm [14,15]. This technique offers excellent potential for obtaining statistically meaningful information with a very high elemental sensitivity of Li, as it is an electropositive element with a high ionisation probability. However, the application of SIMS instruments is typically limited to a lateral resolution of ~ 1 µm [1–7], which is insufficient for revealing the details of Li distribution at the nanoscale.

The development of high-spatial-resolution SIMS instruments in recent years has facilitated imaging of Li distribution with a lateral resolution in the sub-micron range.

For instance, researchers using the NanoSIMS 50L (Cameca, France), using a $Cs^+$ primary ion beam with an impact energy of 16 keV and a lateral resolution of ~ 100 nm, were able to perform nanoscale analysis of Li in $LiCoO_2$-based battery cathode materials by mapping $^7Li^{16}O^-$ to investigate the elemental distribution and the change of surface chemistry during electrochemical cycling [8]. A recently-developed compact design of time-of-flight (ToF) SIMS analyser (TOFWERK, Switzerland) added to a focused ion beam-scanning electron microscope (FIB-SEM) was also used to resolve the segregation of Li on grain boundaries and interfaces of particles in the $LiNi_xMn_yCo_{1-x-y}O_2$ cathodes with a lateral resolution of ~ 50 nm [9,16]. The use of a Zeiss ORION NanoFab Helium/Neon ion microscope equipped with a 25 keV $He^+$ primary ion source and a magnetic mass spectrometer has achieved a lateral resolution of 10 nm by mapping $^7Li^+$ in a lithium titanate and magnesium oxide nanoparticle mixture [17,18].

In addition, SIMS has been demonstrated as a potential method for characterising the distribution of nanoscale Li-rich precipitates in Al-Li alloys. However, most of the existing analyses are limited to a lateral resolution of 1 c 10 μm [4–7] and were not able to spatially resolve the nanoscale phases. An early attempt was made utilising a laboratory-based SIMS instrument developed at the University of Chicago in collaboration with Hughes Research Laboratories (UC-HRL SIM, USA) to spatially resolve the nanoscale δ precipitates that are ~ 100 nm in diameter in an Al-10.7 at. % Li alloy [19,20]. However, the practical implementation of modern commercialised high-spatial-resolution SIMS instruments for nanoscale chemical mapping of Li is rarely reported for Al-Li alloys.

The main aim of the present study is to investigate the capability of two of the high-spatial-resolution SIMS instruments for performing nanoscale chemical mapping in an Al-Li alloy with a particular focus on Li. The SIMS instruments employed are a NanoSIMS 50L (Cameca, France) system and a $Ga^+$ FIB time-of-flight secondary ion mass spectrometry (ToF-SIMS) instrument that comprises a Thermo Scientific Helios Tomahawk ion column FIB-SEM with a compact ToF-SIMS (TOFWERK, Switzerland) analyser. The results of SIMS analysis are further validated by correlative electron microscopy.

The material investigated in this study was an Al-Li-Cu-Mg alloy. The as-cast material was hot-rolled to a thickness of 50 mm and then solution treated, followed by water quenching to room temperature. The material was then pre-stretched at room temperature, followed by a peak ageing treatment (T8) to facilitate the formation of nanoscale strengthening precipitates including δ′ ($Al_3Li$) and $T_1$ ($Al_2CuLi$) phases. Samples for SIMS analysis were extracted from the rolling plane at the mid-thickness (T/2) position. The samples were prepared by grinding with SiC and water, sequential polishing with 3 μm and 1 μm diamond suspensions and a final chemo-mechanical polishing in a diluted solution of 1:1.5 oxide polishing suspension (OPS) to deionised water to give a surface finish without topographical artefacts.

Nanoscale chemical mapping was performed using NanoSIMS 50L and FIB ToF-SIMS in separate regions. The observations obtained from the SIMS analyses were further validated using correlative electron microscopy. This involved backscattered electron (BSE) imaging of secondary phase particles, grain orientation mapping by electron backscatter diffraction (EBSD) and high-resolution analysis using transmission electron microscopy (TEM) on a plan-view thin-foil specimen prepared using a Thermo Scientific Helios $Xe^+$ plasma FIB-SEM. Details of the procedure used for SIMS analysis and correlative electron microscopy are provided in supporting information.

Figure 1 shows the SIMS chemical maps obtained using NanoSIMS and FIB ToF-SIMS over an area of 35 × 30 μm at a pixel size of 68.4 nm. The correlative BSE micrographs collected from the same regions with a pixel size of 34.2 nm are shown in Figures 1b and 1d.

The SIMS maps collected using NanoSIMS and FIB ToF-SIMS (Figures 1a and 1c) both reveal a clear $^7Li^+$ signal from the needle-shaped $T_1$ phases. Whilst as expected the majority of the constituent intermetallic phases show a lower level of $^7Li^+$ signal than the surrounding matrix. The observation of $T_1$ and constituent intermetallic phases is validated by BSE imaging in the same regions, Figures 1b and 1d. This is also consistent with our prior knowledge of the expected phases based on alloy composition and thermomechanical processing. These phases are clearly visualised with distinctive contrast differential to the surrounding matrix due to the abundance of heavier elements such as Cu [21].

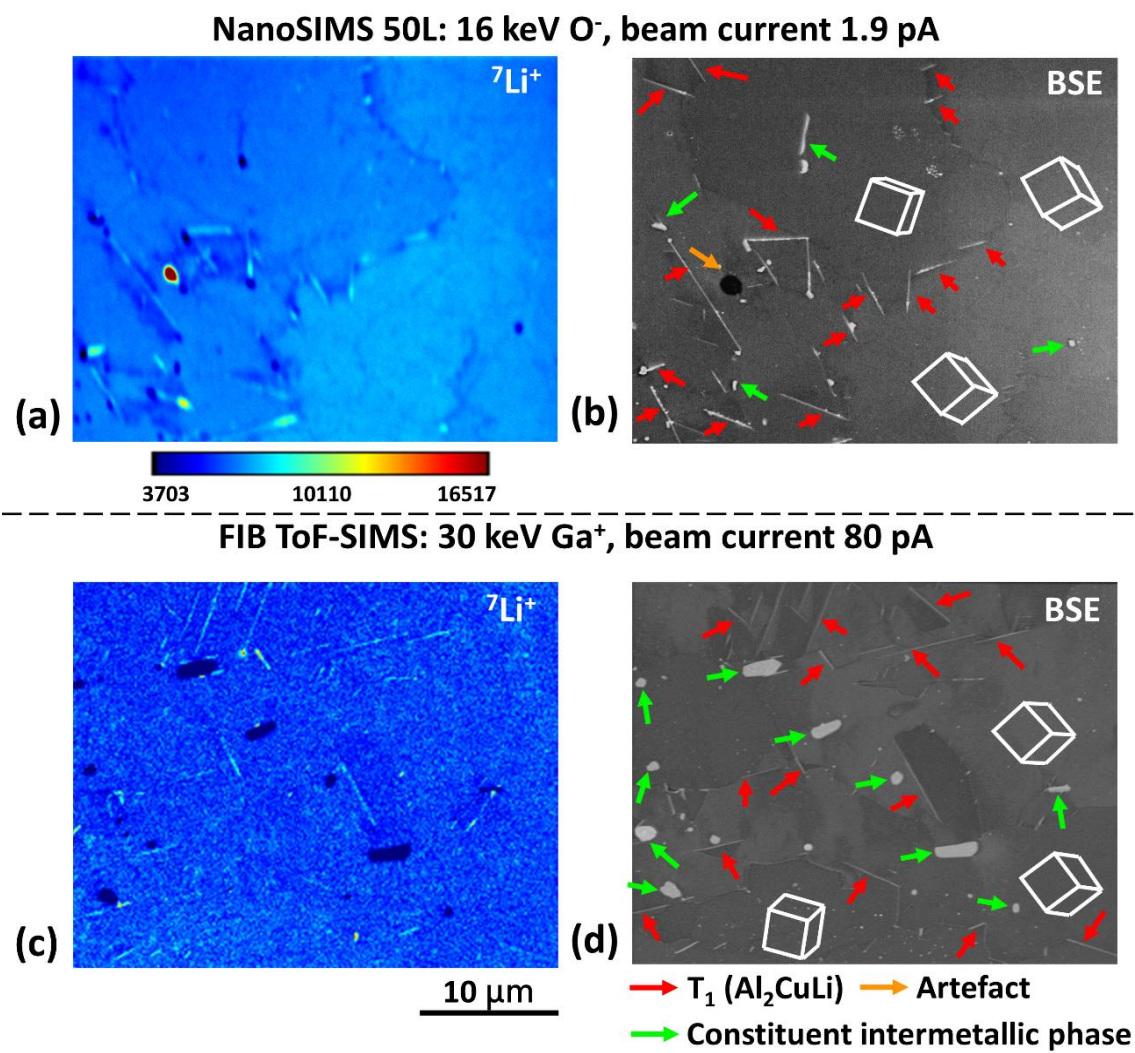

**Figure 1.** (a) NanoSIMS and (c) FIB ToF-SIMS chemical maps showing the distribution of $^7Li^+$ and (b, d) the correlative BSE micrographs obtained after SIMS analysis. The arrows and insets in (b, d) indicate the secondary phase particles present in the area of interest and the crystal orientation measured by EBSD. The colour scale in (a) is presented based on counts per pixel. The count rates of secondary ion signal for both datasets from NanoSIMS and FIB ToF-SIMS are summarised in Table S1.

Figure 2 shows the results of higher resolution SIMS analysis conducted using the FIB ToF-SIMS instrument with a beam current of 24 pA and a pixel size of 29.3 nm. The correlative BSE micrographs collected from the same region with a pixel size of 14.6 nm are shown in Figures 2e and 2f.

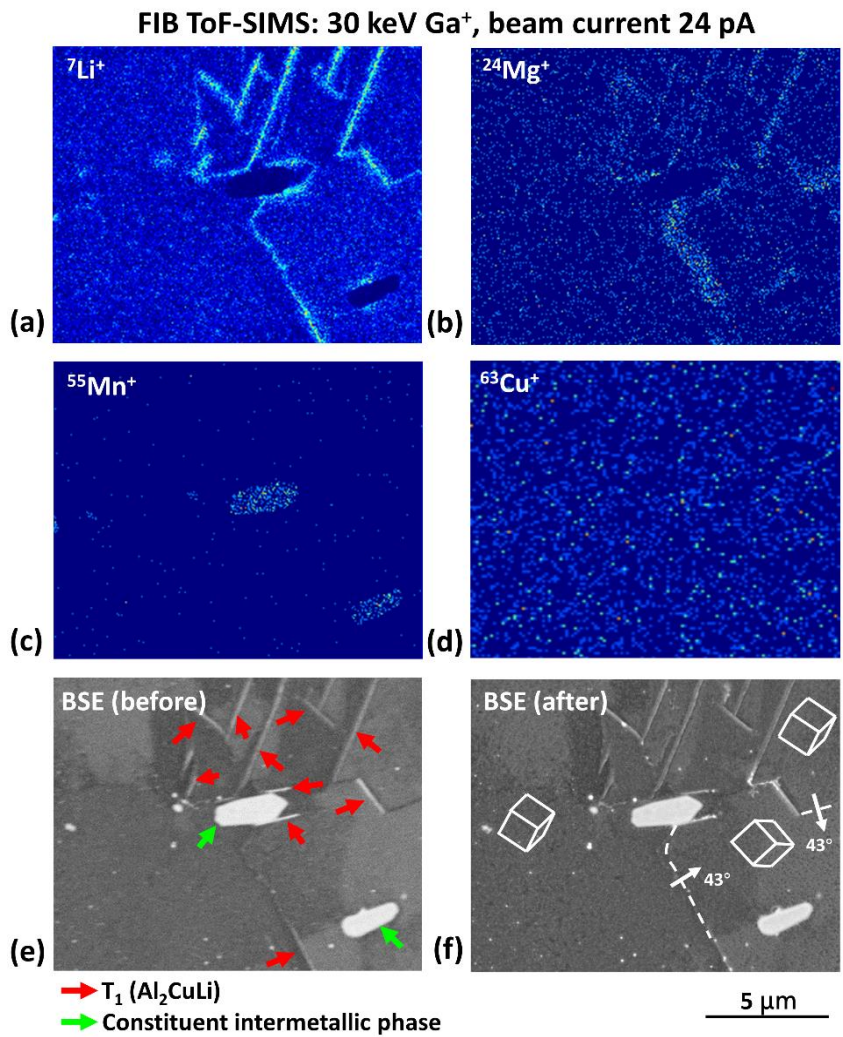

**Figure 2.** (a-d) The FIB ToF-SIMS chemical maps showing the distribution of $^7Li^+$, $^{24}Mg^+$, $^{55}Mn^+$ and $^{63}Cu^+$ and (e, f) the correlative BSE micrographs obtained prior to and after SIMS analysis. The arrows in (e) indicate the secondary phase particles present in the areas of interest. The overlays and arrows in (f) indicate the crystal orientation and misorientation of grain boundaries measured by EBSD, respectively.

The observations of constituent intermetallic phases and the large $T_1$ phases are consistent with the result of FIB ToF-SIMS analysis conducted using a beam current of 80 pA (Figure 1c). In addition, the higher lateral resolution FIB ToF-SIMS chemical maps, Figures 2a and 2b, reveal segregation of Li and Mg on high angle grain boundaries (indicated in Figure 2f by the dashed lines) measuring ~ 90 nm in width. Constituent intermetallic phases exhibit a higher $^{55}Mn^+$ signal than the surrounding matrix, Figure 2c.

Figure 3 shows the $^7Li^+$ SIMS chemical maps and BSE micrographs from magnified regions and the line profiles showing the $^7Li^+$ signal variation across $T_1$ phase particles.

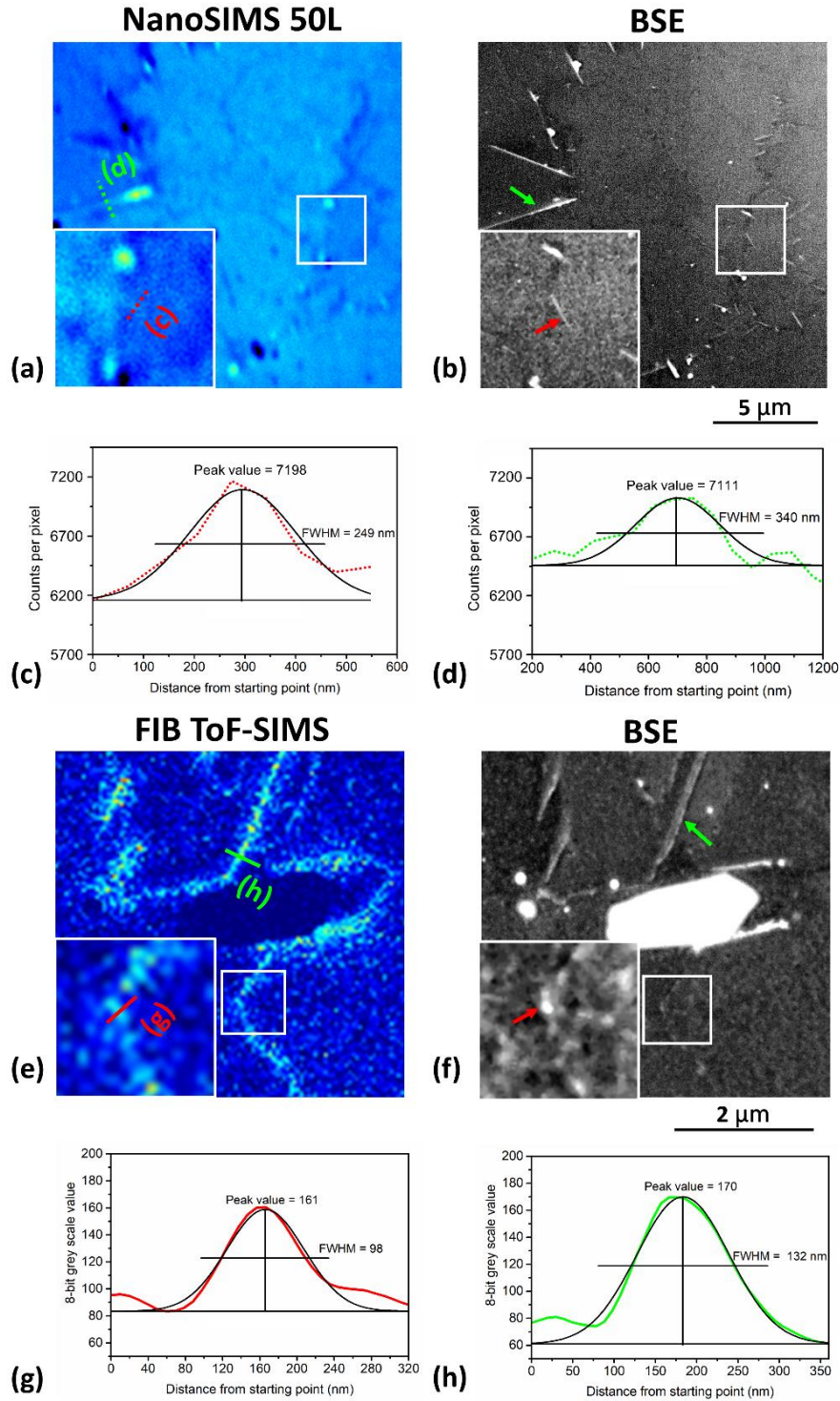

**Figure 3.** (a, e) $^7Li^+$ SIMS chemical maps detailing the distribution of Li within the regions analysed using the NanoSIMS and FIB ToF-SIMS instruments and (b, f) the post SIMS-analysis BSE micrographs from the same regions. The lines in (a, e) indicate the locations where intensity profiles were extracted. The arrows in (b, f) indicate the $T_1$ phase particles examined in detail. The insets

**show the T$_1$ phase particles in magnified views. (c, d, g, h) Graphs showing the intensity profiles across the particles of interest. The black lines show the curves of Gaussian fit as applied using the Origin software for the measurement of full width half maximum (FWHM).**

Intensity profiles with a line thickness of 1 pixel were extracted from the $^7$Li$^+$ chemical maps across the lines shown in Figures 3a and 3e, with the line profiles shown in Figures 3c, 3d, 3g and 3h. The profiles indicate that the widths of the particles measured by NanoSIMS are 249 nm (red) and 340 nm (green) using the full width half maximum (FWHM) method. The lateral resolution of NanoSIMS was further estimated to be 142 nm based on the 16 – 84% criterion (see Figure S1). In the FIB ToF-SIMS image, the particles measure 98 nm (red) and 132 nm (green) using the FWHM method. The lateral resolution of FIB ToF-SIMS was estimated to be 64 nm for the same beam current of 24 pA based on the measurement of lateral resolution from BAM 200L standard (see Figure S2). The T$_1$ phase particles analysed using line profiles were further correlated with the high resolution BSE micrographs from the same regions, Figures 3b and 3f. In Figure 3b the particles indicated by the red and green arrows were clearly visualised and measured ~ 75 nm and ~ 130 nm (FWHM) in width, respectively. In Figure 3f the particles indicated by the red and green arrows measured ~ 45 nm and ~ 120 nm (FWHM) in width, respectively.

The results of FIB ToF-SIMS analysis demonstrating the segregation of Li and Mg have been validated by S/TEM observations of the T$_1$ precipitates at the grain boundary. As demonstrated in Figure S4, the intragranular T$_1$ precipitates that are < 50 nm in length and 1 – 2 nm in width are too small to detect using either SIMS instrument. However, the grain boundary T$_1$ precipitates are formed as clusters that are continuously distributed on the grain boundary, Figure S4a. The clusters measure ~ 90 nm in width, which is consistent with the measured values of widths for the segregated regions showing a strong intensity of $^7$Li$^+$ signal as shown in Figure 2a. In addition, the T$_1$ phase particles measure 200 – 300 nm in length, which is similar to the lengths of Li-rich particles in the FIB ToF-SIMS $^7$Li$^+$ map as shown in Figure 3e. This confirms that the segregation of Li as observed using FIB ToF-SIMS corresponds to the clusters of Li-rich T$_1$ precipitates on the grain boundary.

In summary, the potential of advanced high-spatial-resolution SIMS instruments to achieve chemical mapping with nanoscale lateral resolution has been shown by mapping $^7$Li$^+$ in Al-Li alloys. Li is challenging to detect with many characterisation techniques, but the results presented here show that it is possible to not only map Li but localise it with very high lateral resolution. NanoSIMS analysis was able to resolve features as small as 75 nm in size, whilst FIB ToF-SIMS could visualise the Li in $T_1$ phases as small as 45 nm in size, although neither technique could detect the intragranular precipitates that are smaller in size. The implementation of high-lateral-resolution SIMS instruments for characterising the distribution of Li in modern Al-Li alloys is of significant importance for understanding the influence of microstructure on mechanical and corrosion behaviour. It also offers significant potential for the high-spatial-resolution mapping of Li distribution in Li-ion battery materials.

**Acknowledgement**

The NanoSIMS was funded by UK Research Partnership Investment Funding (UKRPIF) Manchester RPIF Round 2. This work was supported by the Henry Royce Institute for Advanced Materials, funded through EPSRC grants EP/R00661X/1, EP/S019367/1, EP/P025021/1 and EP/P025498/1.